\documentclass[aip,jap,amsmath,amssymb,reprint]{revtex4-1}

\usepackage{graphicx}
\usepackage{dcolumn}
\usepackage{bm}

\usepackage[utf8]{inputenc}
\usepackage[T1]{fontenc}
\usepackage{mathptmx}
\usepackage{hyperref,color}
\usepackage{ulem}
\usepackage{microtype}

\hypersetup{
 pdfnewwindow=true, colorlinks=true,
 linkcolor=blue, anchorcolor=blue,
 citecolor=blue, filecolor=blue,
 menucolor=blue, urlcolor=blue}

\begin{document}

\title{Experimental formation of monolayer group-IV monochalcogenides}

\author{Kai Chang}
\email{changkai@baqis.ac.cn}
\affiliation{Beijing Academy of Quantum Information Sciences, Beijing 100193, China}
\author{Stuart S. P. Parkin}
\email{stuart.parkin@mpi-halle.mpg.de}
\affiliation{Max Planck Institute of Microstructure Physics, Weinberg 2, Halle 06120, Germany}

\date{\today}

\begin{abstract}
Monolayer group-IV monochalcogenides (MX, M = Ge, Sn, Pb; X = S, Se, Te) are a family of novel two-dimensional (2D) materials that have atomic structures closely related to that of the staggered black phosphorus lattice. The structure of most monolayer MX materials exhibits a broken inversion symmetry, and many of them exhibit ferroelectricity with a reversible in-plane electric polarization. A further consequence of the noncentrosymmetric structure is that when coupled with strong spin-orbit coupling, many MX materials are promising for the future applications in non-linear optics, photovoltaics, spintronics and valleytronics. Nevertheless, because of the relatively large exfoliation energy, the creation of monolayer MX materials is not easy, which hinders the integration of these materials into the fast-developing field of 2D material heterostructures. In this Perspective, we review recent developments in experimental routes to the creation of monolayer MX, including molecular beam epitaxy and two-step etching methods. Other approaches that could be used to prepare monolayer MX are also discussed, such as liquid phase exfoliation and solution phase synthesis. A quantitative comparison between these different methods is also presented.
\end{abstract}

\maketitle

\section{Introduction}

\begin{figure*}
\includegraphics[width=\textwidth]{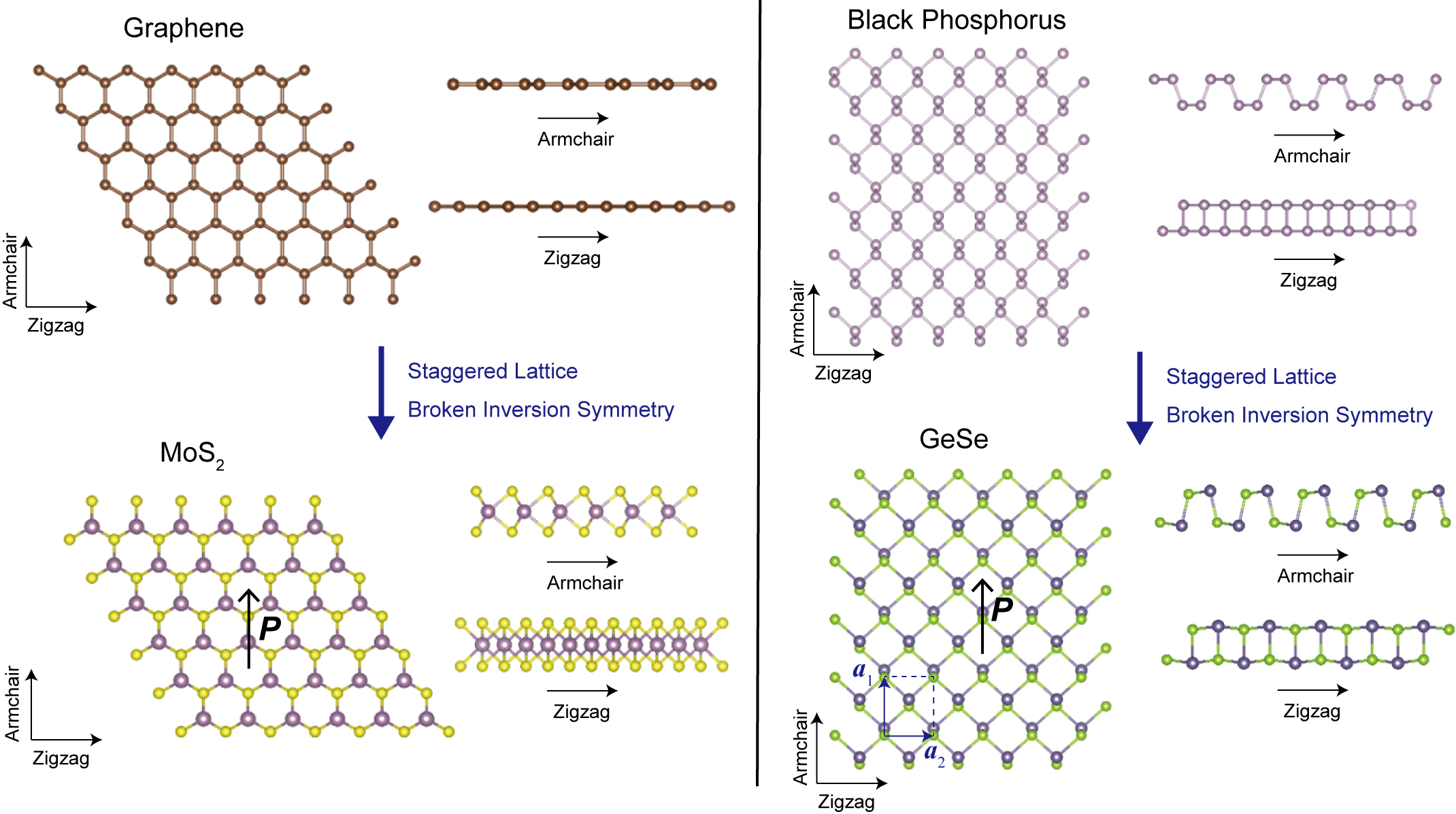}
\caption{Atomic structures of graphene, MoS$_2$, monolayer black phosphorus, and monolayer GeSe (as an example of monolayer MX). For MoS$_2$, the cyan balls correspond to Mo atoms and the yellow balls to S atoms. For GeSe, the silver balls are Ge atoms and the green balls are Se atoms. Similar to the relationship between graphene and the MoS$_2$ lattice, in which the latter can be considered as a staggered graphene lattice without inversion symmetry, the lattice of MX is a staggered version of black phosphorus. Both monolayer MoS$_2$ and MX exhibit in-plane electric polarizations, but in monolayer MX the polarization can be reversed by an external electric field.}\label{MX_structure}
\end{figure*}

Monolayer (ML) group-IV monochalcogenides, abbreviated as MX according to their chemical composition, are a series of 2D semiconductors with many intriguing structural, optical and electronic properties, such as in-plane ferroelectricity, \cite{Chang16_Science_SnTe,Chang19_AM_SnTe,Chang19_APLM_SnTe,Fei15_APL,Mehboudi16_PRL,Fei16_PRL,Wu17_NL,Wang17_2DM} ferroelectric domain wall induced electronic state confinement, \cite{Chang19_PRL_SnTe} second harmonic generation, \cite{Panday17_JPCM,Wang17_NL} photostriction, \cite{Haleoot17_PRL} very large exciton binding energy, \cite{Gomes16_PRB,Luo18_AFM} photovoltaicity, \cite{Rangel17_PRL,Fregoso17_PRB,Panday19_PRB,Wang19_SA} electronic valley polarization and valley Hall effect, \cite{Hanakata16_PRB,Rodin16_PRB,Xu17_PRB} as well as spin splitting of the electronic bands \cite{Gomes15_PRB,Shi15_NL,Rodin16_PRB,Chang19_PRL_SnTe,Liu19_NC,Absor19_PRB} and giant spin Hall effect. \cite{Slawinska19_2DM} Many of these phenomena in MX monolayers originate from their noncentrosymmetric lattice structures, as Figure~\ref{MX_structure} shows. Similar to the monolayers of transition metal dichalcogenides, whose lattice is equivalent to the honeycomb lattice of graphene except that the two sublattices are composed of different atoms, \cite{Xiao12_PRL} MX monolayers have a structure analogus to that of black phosphorus, but with a staggered lattice. In MX monolayers, the broken structural inversion symmetry results in a reversible electric polarization that lifts the valley degeneracy, and spin-orbit coupling simultaneously {lifts} the spin degeneracy. These low-symmetry monolayer materials are especially useful for many non-volatile 2D material heterostructures and devices that have been proposed recently, such as ferroelectric tunneling junctions and corresponding memory devices. \cite{Chang16_Science_SnTe,Shen19_PRApp_FTJ,Shen19_ACSAEM} Recent theoretical developments in this field have been summarized in several review articles. \cite{Wu18_WIRCMS,Cui18_npj2DMA} Nevertheless, compared with the extensive theoretical advances, the development of experiments concerning the physical properties of MX monolayers has evolved much more slowly. \cite{Xia19_NRP,Hu19_Nanotech} This can be attributed, in large part, to the difficulty in creating high-quality monolayer flakes or films. Moreover, the lack of controllable and scalable production methods of MX monolayers also hinders their industrial applications.

\begin{table}
\caption{\label{exf_energy} Comparison of the exfoliation energies of MX materials with other 2D materials.}
\begin{ruledtabular}
\begin{tabular}{ccccc}
Material & $E_{\textrm{exf}}$ (meV/\AA$^2$) & Notes & Reference\footnote{[C] Calculations; [E] Experiments.}\\
\hline
GeS & 36.2 & MX & [C]\cite{Karmodak20_ACSEL} \\
 & 32.5 & & [C]\cite{Lv17_ACBE} \\
 & 28.4 & & [C]\cite{Poudel19_PRM} \\
GeSe & 31.9 & MX & [C]\cite{Karmodak20_ACSEL} \\
 & 28.1 & & [C]\cite{Lv17_ACBE} \\
 & 30.8 & & [C]\cite{Poudel19_PRM} \\
SnS & 34.6 & MX & [C]\cite{Poudel19_PRM} \\
SnSe & 32.0 & MX & [C]\cite{Zhang16_SciRep} \\
 & 55.7 & & [C]\cite{Song18_CPLett} \\
 & 33.6 & & [C]\cite{Poudel19_PRM} \\
Black phosphorus & 29.9 & Isoelectronic of MX & [C]\cite{Zhang16_SciRep} \\
 & 38.2 & & [C]\cite{Karmodak20_ACSEL} \\
Graphite & 11.9 & 2D semimetal & [E]\cite{Liu12_PRB} \\
 & 13.4 & & [E]\cite{Benedict98_CPLett} \\
 & 23.3 & & [E]\cite{Zacharia04_PRB} \\
 & 17.8 & & [C]\cite{Zhang16_SciRep} \\
 & 16.4 & & [C]\cite{Girifalco56_JCP} \\
hBN & 18.2 & 2D insulator & [C]\cite{Zhang16_SciRep} \\
MoS$_2$ & 24.2 & 2D semiconductor & [C]\cite{Zhang16_SciRep} \\
NbSe$_2$ & 23.4 & 2D superconductor & [C]\cite{Karmodak20_ACSEL} \\
In$_2$Se$_3$ & 24.4 & 2D ferroelectric & [C]\cite{Malyi19_Arxiv} \\
WTe$_2$ & 21.7 & 2D ferroelectric & [C]\cite{Karmodak20_ACSEL} \\
MoTe$_2$ & 23.1 & 2D ferroelectric & [C]\cite{Karmodak20_ACSEL} \\
\end{tabular}
\end{ruledtabular}
\end{table}

The difficulty in obtaining MX monolayers is a result of the relatively high exfoliation energy of bulk MX materials. As shown in Table~\ref{exf_energy}, MX materials have significantly higher exfoliation energies \cite{Karmodak20_ACSEL,Lv17_ACBE,Poudel19_PRM,Zhang16_SciRep,Song18_CPLett} than those of other intensively studied 2D materials, such as graphene/graphite, \cite{Liu12_PRB,Benedict98_CPLett,Zhang16_SciRep,Girifalco56_JCP} hexagonal boron nitride (hBN), \cite{Zhang16_SciRep} transition metal dichalcogenides, \cite{Zhang16_SciRep,Karmodak20_ACSEL} and other 2D ferroelectrics such as In$_2$Se$_3$. \cite{Malyi19_Arxiv} Black phosphorus, which is an isoelectronic material to MX, also has a relatively lower exfoliation energy than most of the MX materials. \cite{Zhang16_SciRep}

\begin{table*}
\caption{\label{results_summary} Experimental approaches to create ultrathin MX crystal flakes. \footnote{In this table, one ML refers to one van der Waals layer, which is two atomic layers, or the thicness of 0.5$\sim$0.6 nm.}}
\begin{ruledtabular}
\begin{tabular}{cccccc}
Method & Material & Substrate & Crystal size & Lowest thickness & Reference \\
\hline
Mechanical exfoliation (ME) & GeSe & SiO$_2$/Si & 4$\sim$160 $\mu$m & 60$\sim$140 nm & Mukherjee 2013 \cite{Mukherjee13_GeSe} \\
 & SnSe & SiO$_2$/Si & Tens of $\mu$m & $\sim$100 nm & Tayari 2018\cite{Tayari18_PRB_SnSe} \\
 & SnSe & SiO$_2$/Si & Tens of $\mu$m & 70 nm & Cho 2017\cite{Cho17_NRL_SnSe} \\
Liquid phase exfoliation (LPE) & GeS & - & $\sim$70 nm & 2 MLs & Lam 2018\cite{Lam18_CM_GeS} \\
 & GeSe & - & Up to 200 nm & 4 MLs & Ye 2017\cite{Ye17_CM_GeSe} \\
 & GeSe & - & $\sim$300 nm & 4.3 nm & Ma 2019\cite{Ma19_ACSAMI_GeSe} \\
 & SnS & - & $\sim$100 nm & 4.1 nm & Brent 2015\cite{Brent15_JACS_SnS} \\
 & SnS & - & Up to 180 nm & 2 MLs & Sarkar 2020\cite{Sarkar20_npj2DMA_SnS} \\
 & SnSe & - & $\sim$50 nm & 2 MLs & Huang 2017\cite{Huang17_JPCC_SnSe} \\
LPE + Li ion intercalation & SnSe & - & $\sim$300 nm & 6 MLs & Ju 2016 \cite{Ju16_ACSNano_SnSe,Ju16_CEJ_SnSe} \\
 & SnSe & - & $\sim$1 $\mu$m & 6 nm & Ren 2016\cite{Ren16_MSEB_SnSe} \\
 & SnSe$_{1-x}$S$_x$ & - & Up to 200 nm & 6 MLs & Ju 2017 \cite{Ju17_CM_SnSeS} \\
\hline
Solution phase synthesis & GeS & - & 0.5$\sim$4 $\mu$m & 3$\sim$20 nm & Vaughn 2010 \cite{Vaughn10_JACS_GeS_GeSe} \\
 & GeSe & - & 0.5$\sim$4 $\mu$m & 5$\sim$100 nm & Vaughn 2010 \cite{Vaughn10_JACS_GeS_GeSe} \\
 & SnSe & - & 500 nm & 10$\sim$40 nm & Vaughn 2011 \cite{Vaughn11_ACSnano_SnSe} \\
 & SnSe & - & $\sim$300 nm & 2 MLs & Li 2013\cite{Li13_JACS_SnSe} \\
 & SnS & - & 2$\sim$5 $\mu$m $\times$ 0.5 $\mu$m & 10 nm & Deng 2012\cite{Deng12_ACSNano_SnS} \\
 & SnSe & - & $\sim$500 nm & 3 nm & Zhang 2014 \cite{Zhang14_ACSNano_SnSe} \\
 & SnS & - & 0.1$\sim$1 $\mu$m & 60$\sim$80 nm & Rath 2015 \cite{Rath15_ChemComm_SnS} \\
 & SnS & graphite oxide & $\sim$1 $\mu$m & 4$\sim$5 MLs & Li 2015 \cite{Li15_RSCAdv_SnS} \\
\hline
Molecular beam epitaxy (MBE) & SnTe & Graphene/SiC & Up to 1 $\mu$m & 1 ML & Chang 2016 \cite{Chang16_Science_SnTe}; 2019 \cite{Chang19_APLM_SnTe} \\
 & PbTe & Graphene/SiC & $\sim$300 nm & 1 ML & Chang 2016 \cite{Chang16_Science_SnTe} \\
 & SnSe & Graphene/SiC & $\sim$100 nm & 1 ML & Chang 2020 \cite{Chang20_arxiv_SnSe} \\
Physical vapor deposition (PVD) & GeS & SiO$_2$/Si & Tens of $\mu$m & 30 nm & Li 2012 \cite{Li12_ACSNano_GeS} \\
 & GeSe & Mica & $\sim$5 $\mu$m & 15 nm & Liu 2019 \cite{Liu19_ACSAMI_GeSe} \\
 & SnS & Mica & $\sim$5 $\mu$m & 5.5 nm & Xia 2016 \cite{Xia16_Nanoscale_SnS} \\
 & SnSe & Mica & 1$\sim$6 $\mu$m & 6 nm & Zhao 2015 \cite{Zhao15_NR_SnSe} \\
 & SnSe & PDMS\footnote{Molten polydimethylsiloxane} & 5$\sim$15 $\mu$m & 9$\sim$20 nm & Pei 2016 \cite{Pei16_AEM_SnSe} \\
 & SnS & Au/Si & 10 $\mu$m $\times$ 200 nm & 15 nm & Zhou 2016 \cite{Zhou16_JMCC_SnS} \\
Chemical vapor deposition (CVD) & GeS & SiO$_2$/Si & $\sim$10 $\mu$m\footnote{Width of the GeS nanoribbons.} & 20$\sim$50 nm & Lan 2015 \cite{Lan15_JMCC_GeS} \\
 & SnS & SiO$_2$/Si & $\sim$15 $\mu$m & 139 nm & Yu 2019 \cite{Yu19_JMSME_SnS} \\
 & SnS & SnS$_2$\footnote{Reducing SnS$_2$ flakes into SnS in ethanol vapor.} & 1$\sim$1.5 $\mu$m & 50 nm & Li 2018 \cite{Li18_ASS_SnS} \\
 & SnS & SiO$_2$/Si & 30$\sim$40 $\mu$m & 40$\sim$50 nm & Nalin Mehta 2017\cite{Nalin17_JM_SnS} \\
\hline
CVD + nitrogen etching & SnSe & SiO$_2$/Si & Tens of $\mu$m & 1 ML & Jiang 2017 \cite{2DM17_ML_SnSe} \\
ME + laser etching & GeSe & SiO$_2$/Si & $\sim$2 $\mu$m & 1.5 nm (1 ML?) & Zhao 2018 \cite{Zhao18_AFM_GeSe}; Mao 2018 \cite{Mao18_SR_GeSe} \\
\end{tabular}
\end{ruledtabular}
\end{table*}

During the past two decades, many approaches have been developed to obtain atomically-thin MX flakes, but only a few have reached the limit of a single monolayer. For an overview, Table~\ref{results_summary} {summarizes} the recent progress in creating ultrathin MX single crystalline flakes using various methods.

For experiments concerning 2D materials, mechanical exfoliation is usually the most straightforward approach to create atomically thin flakes, especially in the early stage of material characterization. However, simple mechanical exfoliation experiments using standard scotch tape methods have only yielded MX flakes that are tens of nanometers thick, \cite{Tayari18_PRB_SnSe,Cho17_NRL_SnSe} far from the 2D limit. Using liquid phase exfoliation, during which bulk MX materials are ultrasonicated in various solvents, several-ML thick MX flakes have been obtained. \cite{Lam18_CM_GeS,Ye17_CM_GeSe,Ma19_ACSAMI_GeSe,Brent15_JACS_SnS,Sarkar20_npj2DMA_SnS,Huang17_JPCC_SnSe,Ju16_ACSNano_SnSe,Ju16_CEJ_SnSe,Ren16_MSEB_SnSe,Ju17_CM_SnSeS} The lateral sizes of these MX flakes are typically several hundreds of {nm}, and the lowest thickness so far reached is 2 MLs \cite{Lam18_CM_GeS,Sarkar20_npj2DMA_SnS,Huang17_JPCC_SnSe}. Mechanically separating the last two monolayers for MX materials is especially difficult, probably because of the antiferroelectric coupling between the in-plane polarized monolayers: there is a strong attraction between the edges of the two monolayers that host bound charges with opposite signs. Humidity may also play an important role in the exfoliation process because MX materials can dissolve water on the time scale of just a few nanoseconds even at room temperature. \cite{BL18_ACSCS_WaterSplit}

It is very interesting to reduce the thickness from 2 MLs to a single monolayer because many physical properties of MX change significantly, as these two thicknesses have distinct crystalline symmetries. A monolayer MX flake has broken inversion symmetry because of the in-plane polarization (space group $Pmn2_1$), while 2-ML thick MX restores inversion symmetry because of the antiferroelectric coupling (space group $Pnma$). \cite{Mehboudi16_PRL} For example, spin-splitting of electronic bands exists in monolayer MX flakes, \cite{Gomes15_PRB,Shi15_NL,Rodin16_PRB,Chang19_PRL_SnTe,Liu19_NC,Absor19_PRB} but is absent in the 2-ML thick flakes.

Since it is difficult to obtain monolayer MX flakes through ``top-down'' exfoliation methods, ``bottom-up'' synthesis methods, including solution-phase and vapor-phase routes, have also been extensively studied recently. By controlling the conditions of chemical reactions in solution, one can generate either colloidal MX nanoparticles \cite{Franzman10_SnSe,Baumgardner10_SnSe,Liu11_SnSe_nanowires,Antunez11,Ning11_SnSe_colloidal,Liu14_SnSe_SnS_colloidal} or nanoflakes. \cite{Vaughn10_JACS_GeS_GeSe,Vaughn11_ACSnano_SnSe,Li13_JACS_SnSe,Deng12_ACSNano_SnS,Zhang14_ACSNano_SnSe,Rath15_ChemComm_SnS,Li15_RSCAdv_SnS} Applying a one-pot solution synthesis method, Li \textit{et al.} have created 2-ML thick SnSe nanosheets with lateral sizes of $\sim$300 nm, which is the lowest thickness by far yet achieved through this approach.

Vapor phase synthesis routes include physical vapor deposition (PVD): gaseous MX  molecules are directly deposited onto a {substrate}; chemical vapor deposition (CVD): chemical reactions yielding MX thin flakes happen at the surface of a substrate; and molecular beam epitaxy (MBE): which features ultra-high vacuum deposition through either a physical or chemical process. Usually carried out in tube furnaces, PVD and CVD methods can be carried out at high vapor pressures and create several-nm thick, several-$\mu$m wide MX flakes. \cite{Li12_ACSNano_GeS,Liu19_ACSAMI_GeSe,Xia16_Nanoscale_SnS,Zhao15_NR_SnSe,Pei16_AEM_SnSe,Zhou16_JMCC_SnS,Lan15_JMCC_GeS,Yu19_JMSME_SnS,Li18_ASS_SnS,Nalin17_JM_SnS} Compared with PVD and CVD, MBE has a much stricter requirement as regards the vacuum environment and a much lower deposition rate. In 2016, Chang \textit{et al.} managed to grow the first monolayer MX material --- monolayer SnTe nanoplates --- on graphitized SiC substrates, \cite{Chang16_Science_SnTe} and demonstrated their ferroelectric properties by scanning tunneling microscopy (STM). The lateral size of the monolayer SnTe nanoplates can reach 1 $\mu$m. The experimental growth of monolayer PbTe and SnSe nanoplates was also reported later. \cite{Chang16_Science_SnTe,Chang20_arxiv_SnSe}

It can be seen from Table~\ref{results_summary} that, in general, there is a strong dependence of the lateral crystal size on the thickness of the MX flakes: reducing the thickness typically makes the lateral size smaller. In order to solve this dilemma, some post-etching methods have been developed recently. Jiang \textit{et al.} created 30$\sim$50 $\mu$m large monolayer SnSe flakes by etching CVD grown flakes with nitrogen at elevated temperatures. \cite{2DM17_ML_SnSe} Lasers have also been applied to etch mechanically exfoliated GeSe flakes, yielding 1.5 nm thick, $\mu$m-sized patches. \cite{Zhao18_AFM_GeSe,Mao18_SR_GeSe}

In this focused Perspective, we will give an insight into experimental approaches for creating MX monolayers. As general review articles of the synthesis of MX thin flakes/films have been published elsewhere, \cite{Xia19_NRP,Hu19_Nanotech} the emphasis of this Perspective will be on currently available techniques that can controllably generate MX monolayers. At the same time, we will also briefly review the routes that can create several-ML thick MX flakes, bearing in mind that improved techniques in the future might lead to the successful creation of monolayers. In Section \ref{sec2}, we will review the two existing experimental approaches for obtaining MX monolayers --- MBE and nitrogen post-etching. There are claims that the laser etching method is capable of creating MX monolayers, {but some issues remain to be clarified}: these will also be discussed in this section. In Section \ref{sec3}, we will review other routes that can possibly achieve the monolayer limit in the near future, including liquid phase exfoliation and solution-phase synthesis. We will discuss the advantages and disadvantages of these various approaches, and present an outlook for the future development of this topic in Section \ref{sec4}, and finally conclude the article in Section \ref{sec5}.

Before we begin detailed discussions, it is very important to clearly define the notations that we use here since different notations have been used in the literature. In this Perspective, the orthogonal vectors \textit{\textbf{a}}$_1$ and \textit{\textbf{a}}$_2$ are the crystalline basis along the armchair and zigzag directions of a MX monolayer, as shown in Fig~\ref{MX_structure}. Even the definition of a ``monolayer'' or ``a single layer'' varies in the literature. Here, we shall use the description of ``a monolayer'' (1 ML) to refer to a van der Waals layer, or two atomic layers (AL), which is the smallest possible thickness of MX materials. One ``unit cell'' contains two MLs, or four ALs.

\section{Currently available routes of creating MX monolayers}\label{sec2}

\subsection{Molecular beam epitaxy}

\begin{figure*}
\includegraphics[width=0.9\linewidth]{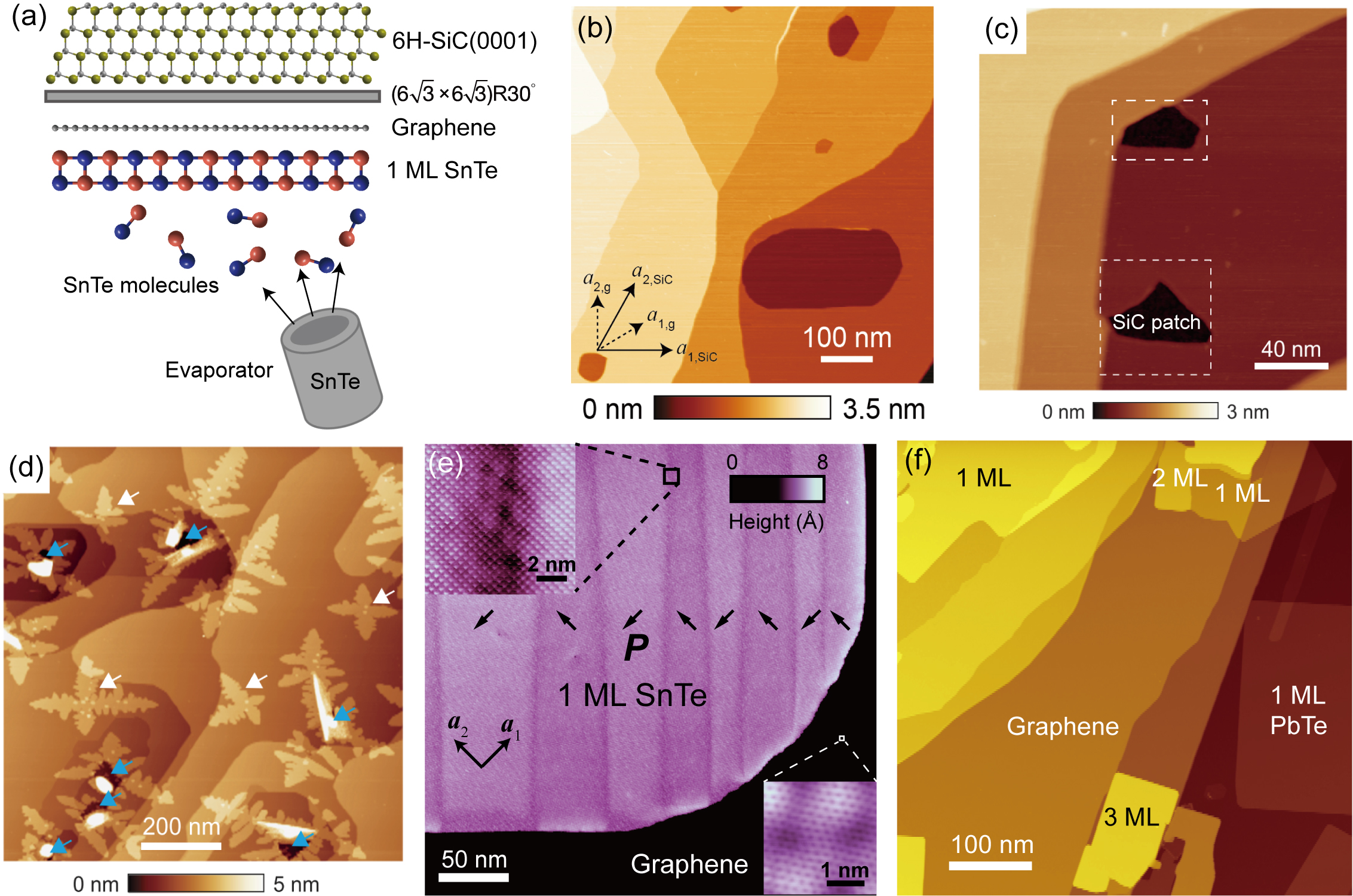}
\caption{(a) Schematic diagram of the MBE growth of monolayer SnTe nanoplates on graphene/SiC substrates. (b)-(f) are STM topography images. (b) A graphene/SiC substrate flash-annealed in ultra-high vacuum. The solid and dashed arrows indicate the crystal basis of SiC and graphene, respectively. (c) Small SiC patches on insufficiently graphitized substrates. (d) The nucleation sites of SnTe at room temperature. The substrate was kept at room temperature during the deposition. All the irregular SnTe islands are one monolayer thick. The cyan arrows label the islands {nucleating} at the SiC patches, and the white arrows indicate those {that} nucleate at the other sites of the surface. Copyright, 2019, authors. \cite{Chang19_APLM_SnTe} Reproduced under CC BY 4.0. (e) Part of a $\sim$1 $\mu$m wide monolayer SnTe nanoplate with 90$^\circ$ domains resolved. Setpoint: sample bias voltage $V_s = -0.2$ V, tunneling current $I_t = 30$ pA. Measured at $T = 4.7$ K. The arrows in each domain indicate the directions of the in-plane polarization. Upper inset, atom-resolved topography image across a domain boundary ($-0.2$ V, 100 pA). Lower inset, atom-resolved topography image of the graphene substrate ($-0.2$ V, 200 pA). (f) PbTe nanoplates grown using similar MBE method as SnTe. Copyright, 2016, American Association for the Advancement of Science. \cite{Chang16_Science_SnTe} Reproduced with permission.}\label{mbe_fig1}
\end{figure*}

In the forty years' history of the MBE growth of MX materials, studies mainly focused on the (001)- and (111)-oriented lead monochalcogenides (PbS, PbSe and PbTe) with rock salt structures, because of their importance as narrow-bandgap semiconductors for mid-infrared lasers and sensors. \cite{SPRINGHOLZ2018211} Most of the films grown for these purposes are thicker than 100 nm. Since the discovery of the topological crystalline insulator (TCI) phase in MX materials such as SnTe, Pb$_{1-x}$Sn$_x$Te and Pb$_{1-x}$Sn$_x$Se, \cite{Fu11_PRL_TCI,Hsieh12_NC_TCI,Dziawa12_NM_TCI,Tanaka12_NP_TCI,Xu12_NC_TCI} and the prediction of a low-dimensional TCI phase in their atomic-thin films, \cite{Liu14_NM_TCI,Ozawa14_PRB_TCI,Liu15_NL_TCI} multiple MBE growth studies of ultrathin MX films have also been reported. \cite{Yan14_PRL_TCI,Yan14_SurfSci_TCI,Polley14_PRB_TCI,Wang15_AM_TCI,Zeljkovic15_NN_TCI,Mandal17_NC_TCI} However, none of these studies have reached the thickness of a single monolayer. Furthermore, all of these films are in rock salt structures, while the MBE growth of staggered black phosphorus structured MX is still rare. \cite{Sutter18_SnS_MBE}

The first MBE experiment that unambiguously reached the monolayer limit of MX was the growth of ferroelectric monolayer SnTe nanosheets on graphitized 6H-SiC(0001) substrates, reported by Chang \textit{et al.} in 2016 \cite{Chang16_Science_SnTe}. (It should be noted that a monolayer of SnTe was referred to as ``one unit cell'' in {that} article, because a rhombic distorted rock salt lattice, whose unit cell contains two atomic layers, was presumed at that time.) In the Supporting Material of this article, the authors also reported the growth of monolayer PbTe nanosheets as an example of a paraelectric MX material for comparison, which has a similar growth mode as SnTe monolayers. In these experiments, SnTe/PbTe nanosheets were deposited from molecular fluxes generated by heating up hBN crucibles filled with high-purity SnTe/PbTe granules in ultra-high vacuum, as illustrated in Fig.~\ref{mbe_fig1}(a). Variable-temperature STM was utilized to characterize the in-plane ferroelectricity in monolayer SnTe nanosheets from the perspectives of lattice distortion, nanosheet edge bound charge induced electronic band bending, and the domain wall motion induced by applying electric field pulses between the STM tip and the sample. Surprisingly, the ferroelectric transition temperature $T_c$ of SnTe monolayers reaches 270 K, much higher than that of bulk SnTe. Subsequent studies revealed the mechanism behind this counterintuitive enhancement phenomenon, which is related to a thickness dependent structural phase transition \cite{Chang19_AM_SnTe,Chang19_APLM_SnTe}. We will further discuss the growth mechanism of SnTe nanosheets in the first section below.

Very recently, monolayer SnSe nanoplates have also been successfully prepared through a two-step MBE growth recipe, also on graphene/SiC substrates. \cite{Chang20_arxiv_SnSe} Controlled room temperature ferroelectric switching and a $T_c$ close to 400 K have been demonstrated in these monolayer nanoplates. In the second section below, we will discuss the growth of monolayer SnSe nanoplates.

\subsubsection{MBE growth of SnTe and PbTe monolayers}

The choice and treatment of substrates is highly important {for} the MBE growth of atomically-thin materials. Compared with other substrates, the graphene surface is extremely smooth and has no dangling bonds, which guarantees a van der Waals epitaxial growth. Since the graphene substrates used for MBE experiments should be uniform and single-crystalline across the surface, monolayer or bilayer graphene, epitaxially grown on the surface of Si-terminated 6H-SiC(0001) substrates were used. There are three approaches to prepare the graphene/SiC substrates: (i) annealing SiC in an H$_2$/Ar gas mixture; \cite{Jia12_CSB_g-SiC} (ii) annealing doped SiC in an ultra-high vacuum environment by passing direct current through the substrate in a Si molecular flux; \cite{Hass08_JPCM_g-SiC} (iii) flash annealing SiC in ultra-high vacuum by direct current heating. \cite{Wang13_JPCM_g-SiC} A full coverage of graphene on the substrate surface is essential for obtaining monolayer nanosheets. As Fig.~\ref{mbe_fig1}(c) shows, on insufficiently graphitized substrates, there are some exposed SiC patches with $(6\sqrt{3}\times 6\sqrt{3})$R30$^\circ$ reconstruction (sometimes also termed a $6\times 6$ reconstruction). \cite{Chang19_APLM_SnTe} Compared with graphene surfaces, these patches have significantly higher surface energy, and as a result, the nucleation rate of SnTe is much higher on these patches, leading to thick clusters hindering the subsequent STM studies.

On properly graphitized SiC substrates, monolayer SnTe nanosheets as large as {$\sim$1} $\mu$m and monolayer PbTe nanosheets of $\sim$300 nm in size have been grown, as shown in Fig.~\ref{mbe_fig1}(e) and (f). \cite{Chang16_Science_SnTe} Specifically, when a monolayer SnTe nanosheet is quickly cooled down below $T_c$, regular parallel or needle-shaped 90$^\circ$ ferroelectric domains are observed by STM. By displaying the hole-like electronic standing waves between parallel domain walls, a space-resolved scanning tunneling spectroscopy study at liquid helium temperature reveals the strong confinement effect of these electrically neutral domain walls to the hole states at the valence band maximum of a SnTe monolayer. \cite{Chang19_PRL_SnTe} The origin of such a strong electronic state modulation effect is attributed to the mismatch of hole valleys in $k$ space across a 90$^\circ$ domain wall, which is a novel type of ``electronic valley quantum well''.

Neither SnTe nor PbTe has a preferred crystalline orientation on graphene, which is another strong indication of van der Waals bonding. At the substrate temperature of 200$^\circ$C, the monolayer nanosheets have nearly rectangular shapes, with atomically smooth edges along the $\langle 11 \rangle$ directions. The nucleation rate increases exponentially as the substrate temperature decreases. \cite{Chang19_APLM_SnTe} However, lowering the substrate temperature also makes the second monolayer easier to form, thus hindering the growth of a fully-capped uniform monolayer film.

\begin{figure*}
\includegraphics[width=\linewidth]{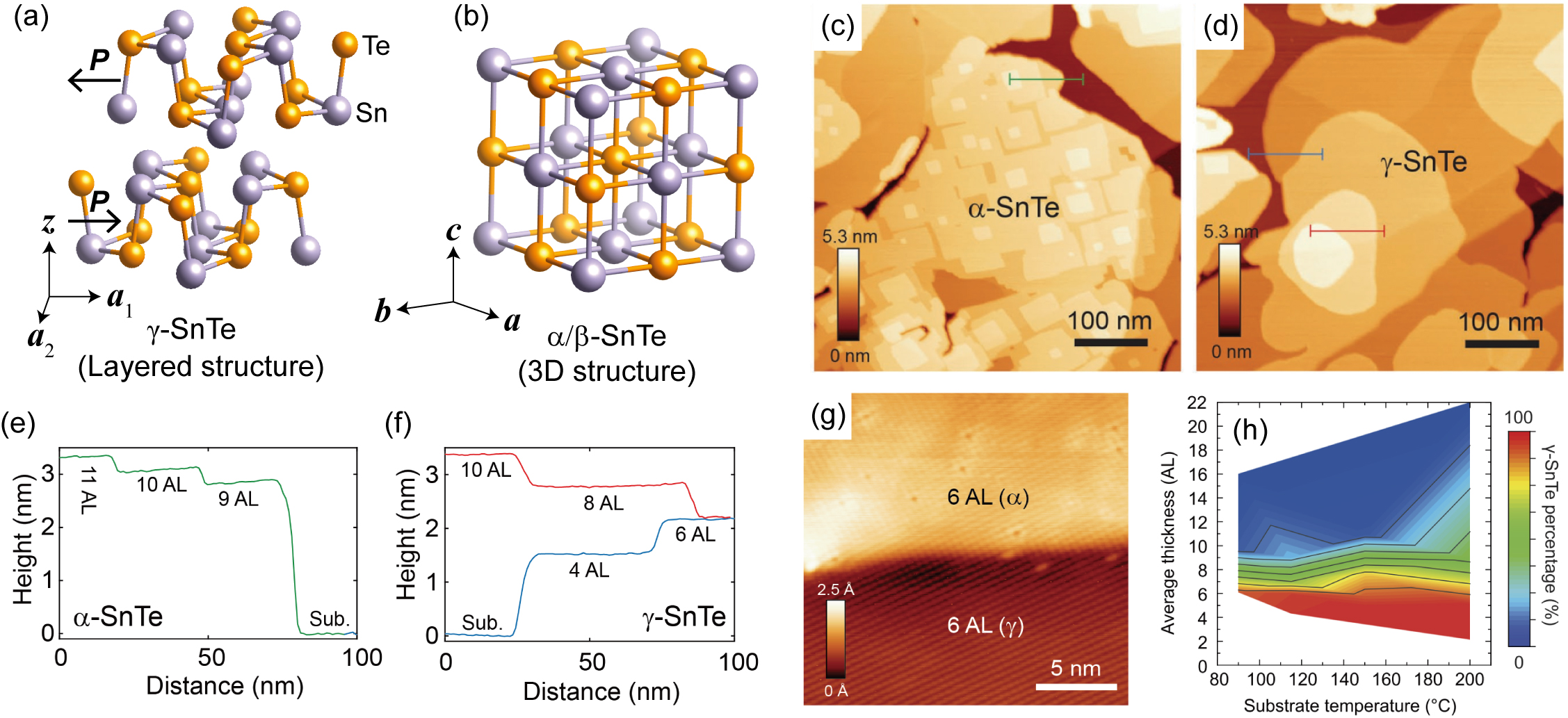}
\caption{(a)-(b) Atomic structures of SnTe in the layered $\gamma$ phase and the 3D $\alpha/\beta$ phase. The $\alpha$ phase has a slight rhombic distortion along (111) from the rock salt structure, and the $\beta$ phase has a undistorted rock salt structure. (c)-(d) STM topography images of the areas in $\alpha$ phase (c) and $\gamma$ phase (d). (e)-(f) Height profiles along the line segments indicated in (c) and (d), respectively. ``AL'' refers to ``atomic layer''. 1 ML contains 2 ALs. (g) Atom-resolved topography image on the boarder of two phases on the same terrace. Setpoint: $V_s = -0.2$ V, $I_t = 30$ pA. All the images were acquired at 77 K. Copyright, 2018, WILEY-VCH Verlag GmbH \& Co. KGaA, Weinheim, reproduced with permission. \cite{Chang19_AM_SnTe} (h) The percentage of $\gamma$-SnTe as a function of the average thickness of the SnTe film and the substrate temperature. Copyright, 2019, authors, reproduced under CC BY 4.0. \cite{Chang19_APLM_SnTe}}\label{mbe_fig2}
\end{figure*}

It is usually impossible to mechanically exfoliate ultrathin SnTe flakes from bulk materials, because it has a three-dimensional (3D) rock salt structure (space group $Fm\overline{3}m$) at room temperature. However, in the aforementioned MBE experiments, the growth of ultrathin SnTe and PbTe nanosheets exhibit strong 2D character. This is the result of a thickness-dependent structural phase transition, as shown in Fig.~\ref{mbe_fig2}(a) and (b). Bulk SnTe has a rock salt structured $\beta$ phase (or a slightly distorted rhombic $\alpha$ phase with the space group of $R3m$ when the temperature is lower than its ferroelectric transition temperature, typically below 100 K), while as the thickness decreases, the layered $\gamma$ phase (space group $Pnma$ for even-ML thick plates and $Pmn2_1$ for odd-ML thick plates), which is isostructural to GeS, GeSe, SnS and SnSe, becomes the most stable structure. \cite{Chang19_AM_SnTe} Besides the lattice structures, $\gamma$-SnTe and $\alpha/\beta$-SnTe are also different in many other respects, as shown in Fig~\ref{mbe_fig2}(c)-(g). The width of atomic terraces can easily reach hundreds of nanometers in $\gamma$-SnTe, while in $\alpha/\beta$-SnTe, this number is only in the tens of nmnanometers. Besides, the height of atomic steps is always 1 ML at the surface of $\gamma$-SnTe, while 0.5-ML (1 atomic layer) atomic steps dominate at the surface of $\alpha/\beta$-SnTe. Last but not least, the concentration of Sn vacancies in $\alpha/\beta$-SnTe ($10^{20}\sim10^{21}$ cm$^{-3}$) is several orders of magnitude higher than that in $\gamma$-SnTe ($10^{17}\sim10^{18}$ cm$^{-3}$), which introduces a large number of p-type carriers and results in a deterioration of ferroelectricity in bulk SnTe. The robust in-plane ferroelectricity in monolayer SnTe nanoplates is found in $\gamma$-SnTe.

All the 1-ML, 2-ML and nearly all the 3-ML thick SnTe nanosheets take up the $\gamma$ phase, while above 4-ML, the ratio of the $\alpha/\beta$ phase gradually increases. There is not a certain critical thickness for the structural phase transition, because the $\gamma$ phase is still metastable above the thickness of 4-ML. Nevertheless, $\gamma$-SnTe nanosheets thicker than 10 MLs {are} very rare, as shown in Fig~\ref{mbe_fig2}(h).

The discovery of a structural phase transition in SnTe nanosheets unambiguously bridges monolayer SnTe with the monolayers of other ferroelectric group-IV monochalcogenides, which have been theoretically predicted since {2013 \cite{Tritsaris13_JAP,Singh14_APL,Mehboudi16_PRL,Fei16_PRL,Wu17_NL,Wang17_2DM}}. Although detailed thickness-dependent study of PbTe nanosheets is still absent, it is very likely that a similar 3D-to-2D structural phase transition also takes place, which accounts for the large monolayer PbTe nanosheets in Fig~\ref{mbe_fig1}(f).

\subsubsection{Two-step MBE growth of SnSe monolayers}

\begin{figure}
\includegraphics[width=\linewidth]{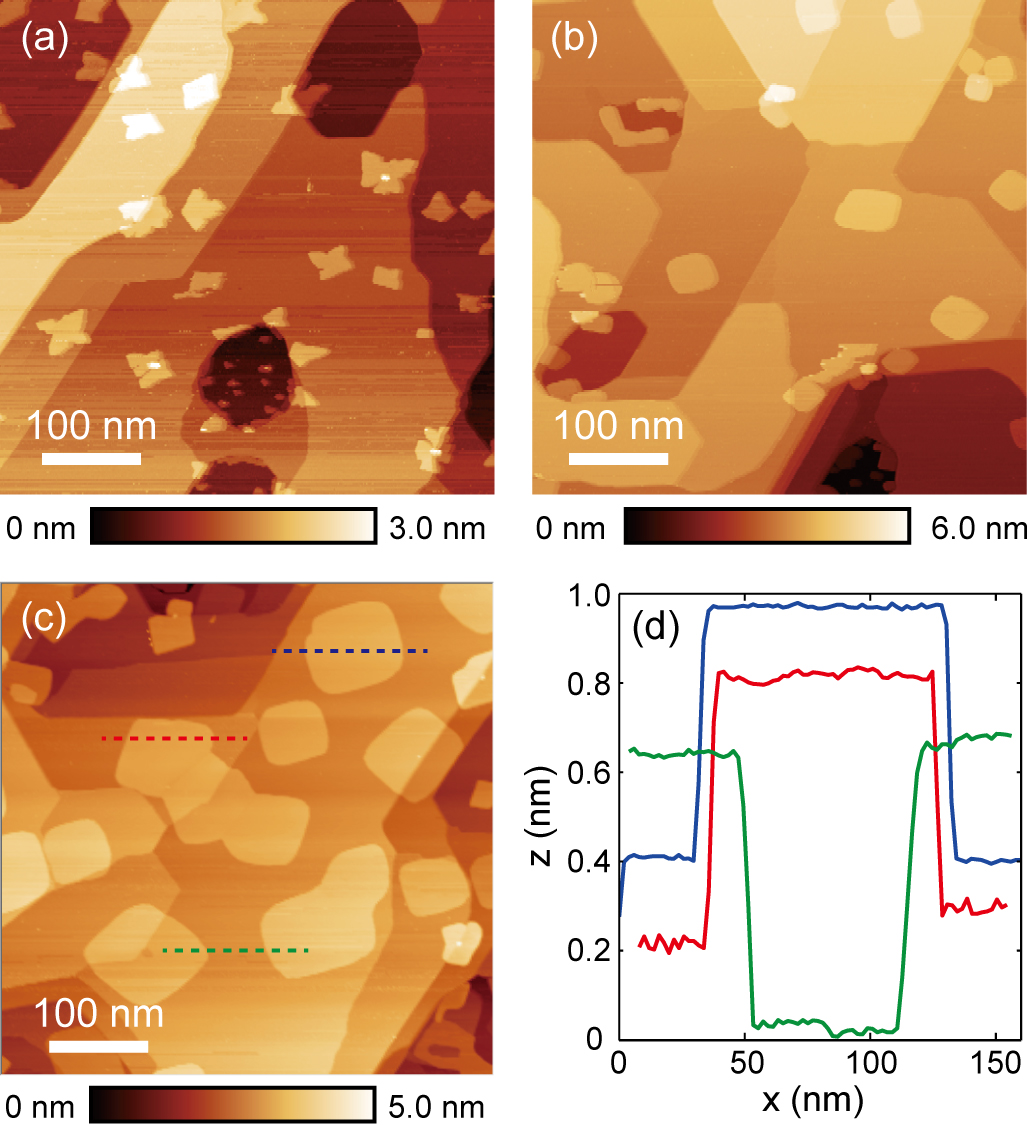}
\caption{(a) STM topography image of the nuclei of monolayer SnSe deposited on a graphene substrate. The substrate was kept at 40$^\circ$C during the 30s long deposition. (b) The same sample as in (a) after annealing at 210$^\circ$C for 1h. (c) The topography after a second deposition lasting for 180s, with the substrate temperature set to 210$^\circ$C. Setpoint: $V_s = -0.3$ V, $I_t = 2$ pA. (d) Height profiles extracted along the dashed lines in (c). The curves are vertically shifted for clarity. All the SnSe plates show similar heights of $\sim$0.6 nm.}\label{mbe_fig3}
\end{figure}

Although bulk SnSe itself has a layered structure, it is nevertheless difficult to grow large-area monolayer SnSe nanoplates on graphene/SiC substrates with a single deposition process, probably because of a higher surface energy in SnSe, induced by a larger lattice buckling compared with that of $\gamma$-SnTe. \cite{Mehboudi16_PRL,Chang19_AM_SnTe} The dilemma in the one-step MBE growth of SnSe is that, at relatively low substrate temperatures (70$^\circ$C, for instance), the nuclei are 1-ML thick, but the second monolayer starts to grow when the nanoplates are only tens of nanometers wide; while at higher substrate temperatures, the nuclei are thicker than 1 ML. A temperature window that can balance the nuclei thickness and nanoplate size has not {been found yet}.

Very recently, a two-step MBE growth recipe has been developed to prepare monolayer SnSe nanoplates with widths above 100 nm. \cite{Chang20_arxiv_SnSe} The procedure is demonstrated in Fig.~\ref{mbe_fig3}. First, the substrate temperature is stabilized at 40$\sim$50$^\circ$C during a first deposition that lasts for only 30 seconds, so that the second monolayer of SnSe does not grow. Then the sample is annealed at 210$^\circ$C for 1 hour, {time during which} the irregular SnSe islands become rectangular. Finally, a second deposition is carried out using the same parameters as the first one, except that the substrate temperature is kept at $210^\circ$C. The small monolayer SnSe islands grow into larger nanoplates which retain a rectangular shape.

Interestingly, because of the coincidental lattice matching between graphene ($a_g=2.46$~\AA) and the $a_2$ of monolayer SnSe (4.26~\AA, which is $\sqrt{3}$ times as large as $a_g$), the monolayer SnSe nanoplates follow a highly oriented growth mode, which is very different from monolayer SnTe and PbTe. According to both reflection high energy electron diffraction (RHEED) patterns and atom resolved topography images recorded by STM, the \textit{\textbf{a}}$_1$ direction of SnSe is always parallel to the zigzag direction of graphene. \cite{Chang20_arxiv_SnSe} The oriented growth of SnSe nanoplates is greatly beneficial {for} anisotropic characterization experiments, such as angle-resolved photoemission spectroscopy (ARPES) and second-harmonic generation (SHG) measurements.

It should be noted that the phase diagram of SnSe \cite{Sharma86_PD_SnSe} has a significant difference compared with those of either SnTe \cite{Sharma86_PD_SnTe} or PbTe. \cite{Lin89_PD_PbTe} Sn (Pb) and Te form only one stable binary compound, which is Sn(Pb)Te; however, Sn and Se can form both SnSe and SnSe$_2$. Therefore, when growing SnTe (PbTe) from separate Sn (Pb) and Te evaporation sources, it is safe to apply extra Te flux, as the excessive Te molecules will re-evaporate at sufficiently high substrate temperatures. Nevertheless, if SnSe is grown in this way, the Se flux must be carefully controlled to prevent the formation of SnSe$_2$. Another simpler solution is to directly evaporate SnSe molecules from high-purity SnSe granules, which is the method adopted in the two-step growth process described above.

In principle, other MX monolayers, such as GeS, GeSe and SnS, can also be grown in a similar way as SnSe, if their nuclei are also 1-ML thick at certain temperatures. However, given that these materials have larger lattice buckling than SnSe, the size of nuclei islands in the first step probably needs to be further reduced. Currently, the lateral size of SnSe nanoplates prepared by this method is mainly limited by the distance between neighboring nuclei. Controlling the density of nucleation centers on graphene substrates is an important topic for further studies.

\subsection{Nitrogen post-etching method}

\begin{figure*}
\includegraphics[width=\textwidth]{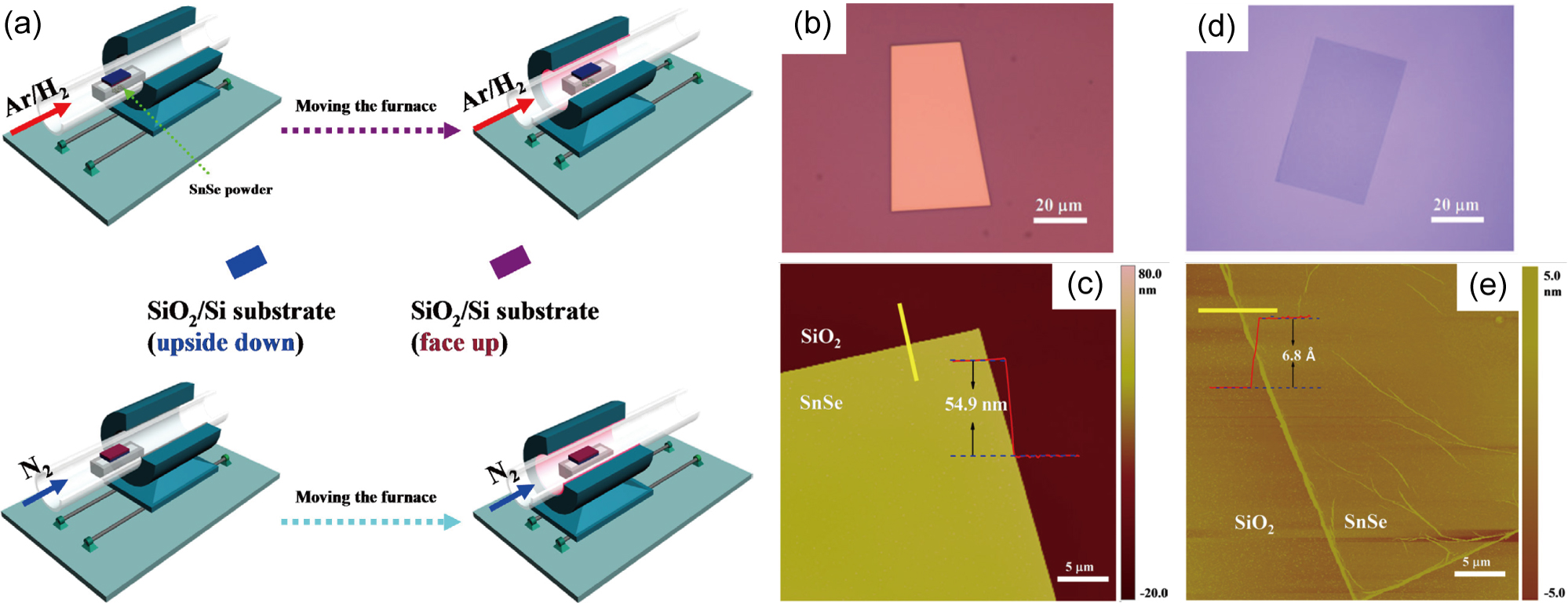}
\caption{(a) Schematic diagrams of the growth of bulk rectangular SnSe flakes by vapor transport deposition (upper panels) and monolayer rectangular SnSe flakes resulting from nitrogen etching (lower panels). (b) Optical image of an as-synthesized bulk rectangularly shaped SnSe flake. (c) A typical atomic force microscopy (AFM) image at the edge of a bulk flake. (d) Optical image of an as-synthesized monolayer SnSe flake. (e) A typical AFM image at the edge of a monolayer SnSe flake. Copyright, 2017, IOP Publishing Ltd. \cite{2DM17_ML_SnSe} Reproduced with permission.}\label{cvd-etching}
\end{figure*}

In 2017, Jiang \textit{et al.} reported a two-step growth-etching method of obtaining monolayer SnSe flakes that are tens of $\mu$m wide, as shown in Fig.~\ref{cvd-etching}. \cite{2DM17_ML_SnSe} The first step is the growth of rectangular SnSe flakes by vapor transport deposition, which is relatively routine. The SnSe powders, placed within a ceramic boat, are evaporated at 700$^\circ$C in a tube furnace in an Ar/H$_2$ flow: this results in $\sim$50 nm thick rectangularly shaped SnSe flakes to be deposited onto a SiO$_2$/Si substrate that is upside down, facing the boat. In a second step, a nitrogen flow is introduced into the tube furnace, and the SnSe flakes, thereby, are etched at 700$^\circ$C, during which the rectangular shape of the flakes are retained but their thickness is reduced to a single monolayer. Electrical transport experiments on these monolayer SnSe flakes show that they are p-type intrinsically doped, which is consistent with the MBE grown monolayer SnSe nanoplates that was described in the previous section.

The mechanism behind this etching process is intriguing. The authors have tried different types of etching gases, including pure Ar, an Ar/H$_2$ mixture, and pure H$_2$. The etching effect of pure Ar is very weak, while pure H$_2$ causes severe deterioration at the surface of the flakes. An Ar/H$_2$ mixture has a similar etching effect as nitrogen, but the latter is safer since it is not flammable. Furthermore, according to the experiments in which the etching times were varied from 1 to 20 min, there is, likely, a self-limiting mechanism: the etching halts once the whole flake is uniformly 1-ML thick. This mechanism is especially ideal for the fabrication of devices based on large monolayer SnSe flakes. The reason for this self-limiting etching process is still not clear. Nevertheless, the etching process is not layer-by-layer, but rather starts from the edges of a flake, which hinders precise thickness control for creating uniform several-ML thick flakes.

\subsection{Laser post-etching method}

\begin{figure}
\includegraphics[width=\linewidth]{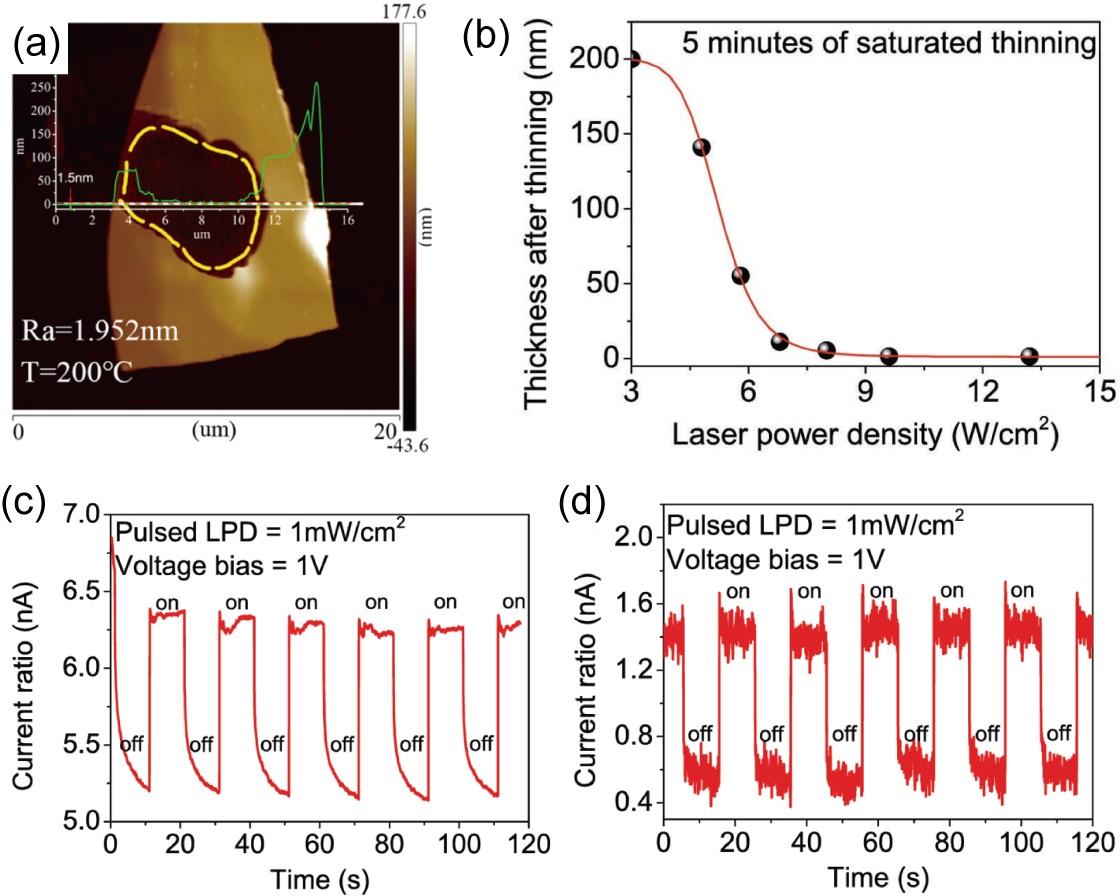}
\caption{(a) AFM topography image of a mechanically exfoliated GeSe nanosheet, with the central area laser etched down to a thickness of 1.5 nm. The sample is subsequently annealed at 200$^\circ$C to further decrease the roughness of the etched area. Copyright, 2018, authors. \cite{Mao18_SR_GeSe} Reproduced under CC BY 4.0. (b) Thinning laser power densities dependence of the average minimum layer thickness. (c) and (d) Photoresponse characteristics of the devices made from a pristine GeSe nanosheet (c) and a second one that is laser thinned to 1.5 nm (d). Copyright, 2018, WILEY-VCH Verlag GmbH \& Co. KGaA, Weinheim. \cite{Zhao18_AFM_GeSe} Reproduced with permission.}\label{laser}
\end{figure}

Etching using a high-power laser is another approach that can etch the mechanically exfoliated MX flakes down to atomic thicknesses. In 2018, {Zhao \textit{et al.} and Mao \textit{et al.}} reported the laser etching of GeSe flakes, from over 100 nm to a lowest thickness of 1.5 nm, as illustrated in Fig.~\ref{laser}(a) and (b). \cite{Zhao18_AFM_GeSe,Mao18_SR_GeSe} The lateral size of the thinned areas can reach several $\mu$m in extent. Similar to the nitrogen etching discussed above, this laser etching procedure also shows a self-limiting behavior. As the laser power density is gradually increased to $9.6\times 10^4$ W/cm$^2$ (with an etching time fixed at 5 min), the minimum thickness of the thinned area decreases monotonically to 1.5 nm; while when the laser power density is further increased, the minimum thickness stops decreasing. This property makes such a laser thinning method a robust approach for fabricating ultrathin MX flakes with a certain thickness. The authors reported an optimal thinning laser power density of $36.7\times 10^4$ W/cm$^2$, at which the resulting thinned GeSe exhibits maximum photoluminescence intensity. Furthermore, as compared with bulk GeSe, the 1.5-nm thick GeSe exhibits a faster response and higher sensitivity in photocurrent measurements [see Fig.~\ref{laser}(c) and (d)], implying an indirect-to-direct band gap transition as the thickness is decreased, and suggesting that atomically-thin GeSe is a promising photodetection and photovoltaic material.

However, it should be noted that, although the authors ascribe the thinnest GeSe (1.5 nm) to a single monolayer, this thickness is not consistent with other reports: both the MBE grown and nitrogen etched monolayer MX flakes reviewed above show thicknesses of $0.5\sim$0.6 nm in STM or AFM measurements, and the other 2-ML thick MX flakes that will be discussed below are 1.0$\sim$1.1 nm thick from AFM studies. Further studies are needed to clarifying this thickness mystery and the detailed mechanisms of laser etching.

\section{Other possible routes to create monolayer MX}\label{sec3}

\subsection{Liquid phase exfoliation}

\begin{figure*}
\includegraphics[width=\linewidth]{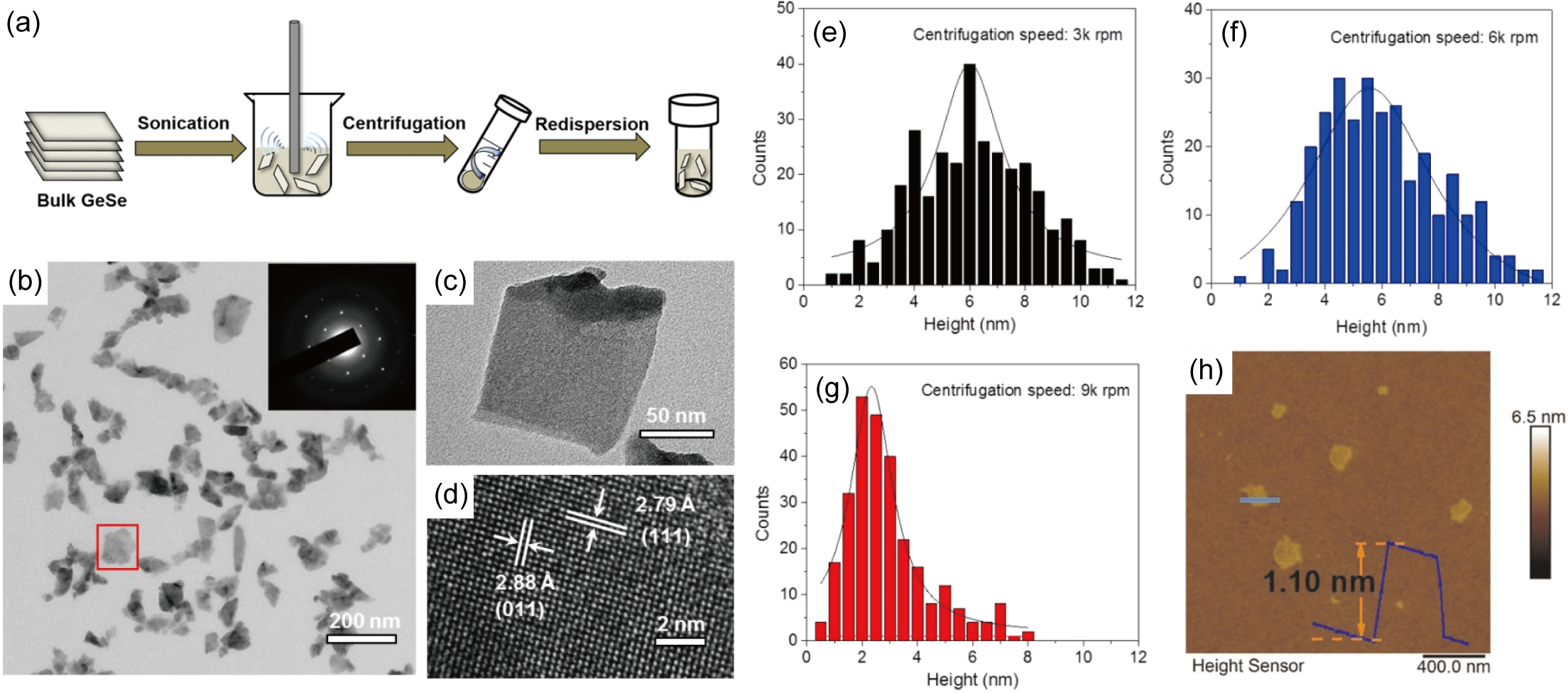}
\caption{(a) Schematic illustration of liquid phase exfoliation of MX materials. (b) Transmission electron micrography (TEM) image of exfoliated GeSe sheets. Inset, selected area electron diffraction (SAED) pattern of the nanosheet inside the red rectangle. (c) TEM image of a single GeSe nanosheet. (d) High-resolution TEM image of GeSe nanosheet with lattice fringes. (e)-(g) Statistics of the thicknesses of GeSe nanosheets collected at different centrifugation speeds. Copyright, 2017, American Chemical Society. \cite{Ye17_CM_GeSe} Reproduced with permission. (h) AFM image of exfoliated SnS nanosheets. The height profile is from a 2-ML thick nanosheet as indicated by the horizontal line. Copyright, 2020, authors. \cite{Sarkar20_npj2DMA_SnS} Reproduced under CC BY 4.0.}\label{lpe}
\end{figure*}

Since a simple scotch tape exfoliation method has not yet been shown to generate atomically-thin MX flakes, liquid phase exfoliation recipes have been developed, which include three main steps, as illustrated in Fig~\ref{lpe} (a). In a first step, bulk MX crystalline granules or powders are ultrasonicated in a solvent, during which the cavity bubbles generated by the intense sound waves will collapse to form a high energy jet, which breaks the bulk layered compounds into thin sheets. \cite{Ye17_CM_GeSe} Various solvents have been used in different studies, including water, hexane, ethanol, acetone, chloroform, N-methylpyrrolidone (NMP), dimethylformamide (DMF), and isopropyl alcohol (IPA), \textit{etc.} \cite{Ju16_ACSNano_SnSe,Ju16_ACSNano_SnSe,Ju17_CM_SnSeS,Lam18_CM_GeS,Ye17_CM_GeSe,Ma19_ACSAMI_GeSe,Brent15_JACS_SnS,Sarkar20_npj2DMA_SnS,Huang17_JPCC_SnSe,Ren16_MSEB_SnSe} Then, the solvent is centrifuged to separate out the larger particles, leaving only nanosheets. Finally, the resulting nanosheets can be redistributed in another solvent to form a stable colloidal dispersion. Fig.~\ref{lpe}(b)-(d) show typical TEM images of the nanosheets obtained via such a liquid phase exfoliation approach. The lowest thickness of MX nanosheets achieved in this approach is 2 MLs, with lateral sizes ranging from several tens of nanometers to $\sim$200 nm.

Several factors are important in determining the yield from liquid phase exfoliation procedures. First, the surface tension and polarity of the solvents, which influence the rate of re-stacking of the nanosheets, play essential roles in the exfoliation processes. \cite{Coleman11_Science_LPE} A comparative study suggests that NMP yields the darkest dispersion of GeS, implying a high exfoliation efficiency and stable dispersion. \cite{Lam18_CM_GeS} Second, a higher centrifugation speed tends to yield thinner nanosheets, as illustrated in Fig~\ref{lpe}(e)-(g). With a centrifugation speed of 9000 rpm, the peak of the resulting thickness distribution curve can be pushed to 2 nm, corresponding to 4 MLs.

Furthermore, given that MX materials are promising for lithium-ion batteries because of their high energy capacity, \cite{Cho13_CC_LiBatt} high Li$^+$ diffusion coefficient \cite{Zhou16_JMCA_LiBatt} and low diffusion barrier, \cite{Li16_JMCA_LiBatt} there have been a series of studies concerning liquid phase exfoliation after Li ion intercalation into bulk MX materials, \cite{Ju16_ACSNano_SnSe,Ju16_CEJ_SnSe,Ren16_MSEB_SnSe,Ju17_CM_SnSeS} which are also very useful for photovoltaic and thermoelectric devices.

\subsection{Solution phase synthesis}

\begin{figure}
\includegraphics[width=\linewidth]{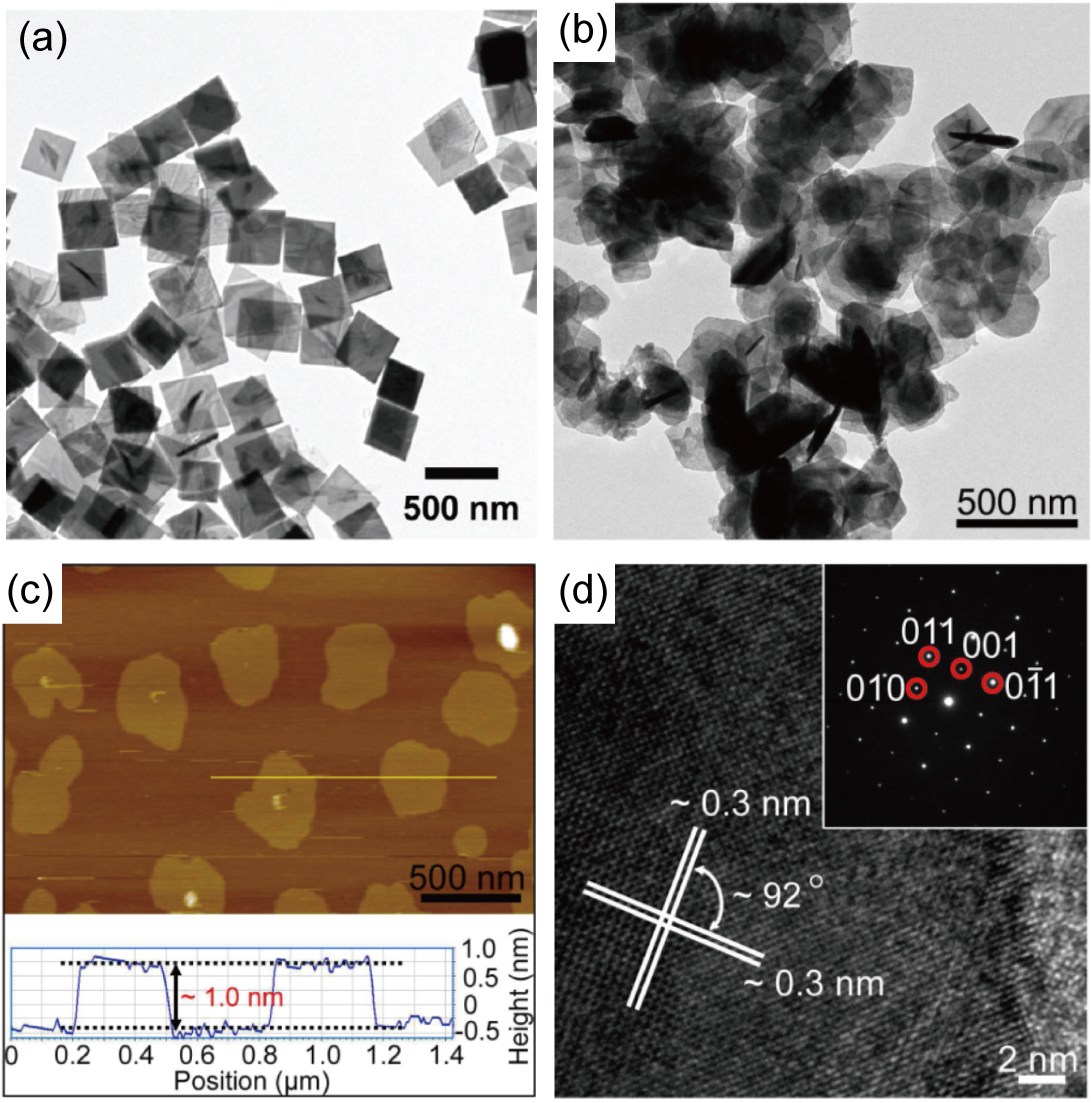}
\caption{SnSe nanosheets synthesized with colloidal one-pot reaction methods. (a) TEM image of rectangularly shaped SnSe nanosheets with thicknesses ranging from 10 to 40 nm. Copyright, 2011, American Chemical Society. \cite{Vaughn11_ACSnano_SnSe} Reproduced with permission. (b) TEM and (c) AFM images of SnSe nanosheets with thickness of 2 MLs (one unit cell, or 4 atomic layers). (d) HRTEM image and SAED pattern of a SnSe nanosheet. Copyright, 2013, American Chemical Society. \cite{Li13_JACS_SnSe} Reproduced with permission.}\label{one_pot}
\end{figure}

As a fast and low-cost method for producing nanomaterials for large scale industrial applications, the solution phase synthesis of MX nanosheets has drawn significant attention. Because of the highly anisotropic crystalline structure of MX materials, colloidal thin flakes can be generated through chemical reactions that take place in organic solvents \cite{Vaughn10_JACS_GeS_GeSe,Vaughn11_ACSnano_SnSe,Li13_JACS_SnSe,Deng12_ACSNano_SnS,Zhang14_ACSNano_SnSe,Rath15_ChemComm_SnS,Li15_RSCAdv_SnS}.
All of these studies use inorganic halides as the source of the group-IV elements, such as SnCl$_4$, \cite{Li13_JACS_SnSe} SnCl$_2$, \cite{Vaughn11_ACSnano_SnSe,Zhang14_ACSNano_SnSe,Rath15_ChemComm_SnS,Li15_RSCAdv_SnS} GeI$_4$, \cite{Vaughn10_JACS_GeS_GeSe} SnI$_4$, \cite{Deng12_ACSNano_SnS} \textit{etc.} The source of the group-VI elements is usually organic material, such as dodecanethiol, \cite{Vaughn10_JACS_GeS_GeSe} trioctylphosphine selenide
(TOP-Se), \cite{Vaughn10_JACS_GeS_GeSe,Vaughn11_ACSnano_SnSe,Zhang14_ACSNano_SnSe} and thioacetamide, \cite{Rath15_ChemComm_SnS} while recipes with cheaper and/or less toxic inorganic materials such as NaHS, \cite{Deng12_ACSNano_SnS} SeO$_2$ \cite{Li13_JACS_SnSe} and (NH$_3$)$_2$S \cite{Li15_RSCAdv_SnS} have also been developed. These reactants are mixed together with organic solvents (for example, oleylamine with phenanthroline dissolved in it) \cite{Li13_JACS_SnSe} in a flask, heated up to reaction temperatures ranging from 180$^\circ$C to 320$^\circ$C, then the products are dispersed into new organic solvents, and centrifuged to separate away large particles. The last two steps are repeated until stable colloidal nanosheets are obtained.

With this procedure, one can usually obtain several-nm-thick nanosheets with lateral sizes of several hundred nanometers. Specifically, in 2013, Li \textit{et al.} reported the synthesis of $\sim$300 nm-wide and 2-ML thick SnSe nanosheets (referred to as ``single-layer'' in the article, which actually means a single unit cell, or 4 atomic layers) using a one-pot recipe, which was the thinnest MX nanosheet that had been formed at that time [Fig.~\ref{one_pot}(b)-(d)]. \cite{Li13_JACS_SnSe}

Substrates can also be applied in solution-phase synthesis. For example, Li \textit{et al.} reported the synthesis of SnSe nanosheets on graphite oxide surfaces in 2015. \cite{Li15_RSCAdv_SnS}

\section{Discussion}\label{sec4}

\begin{figure}
\includegraphics[width=\linewidth]{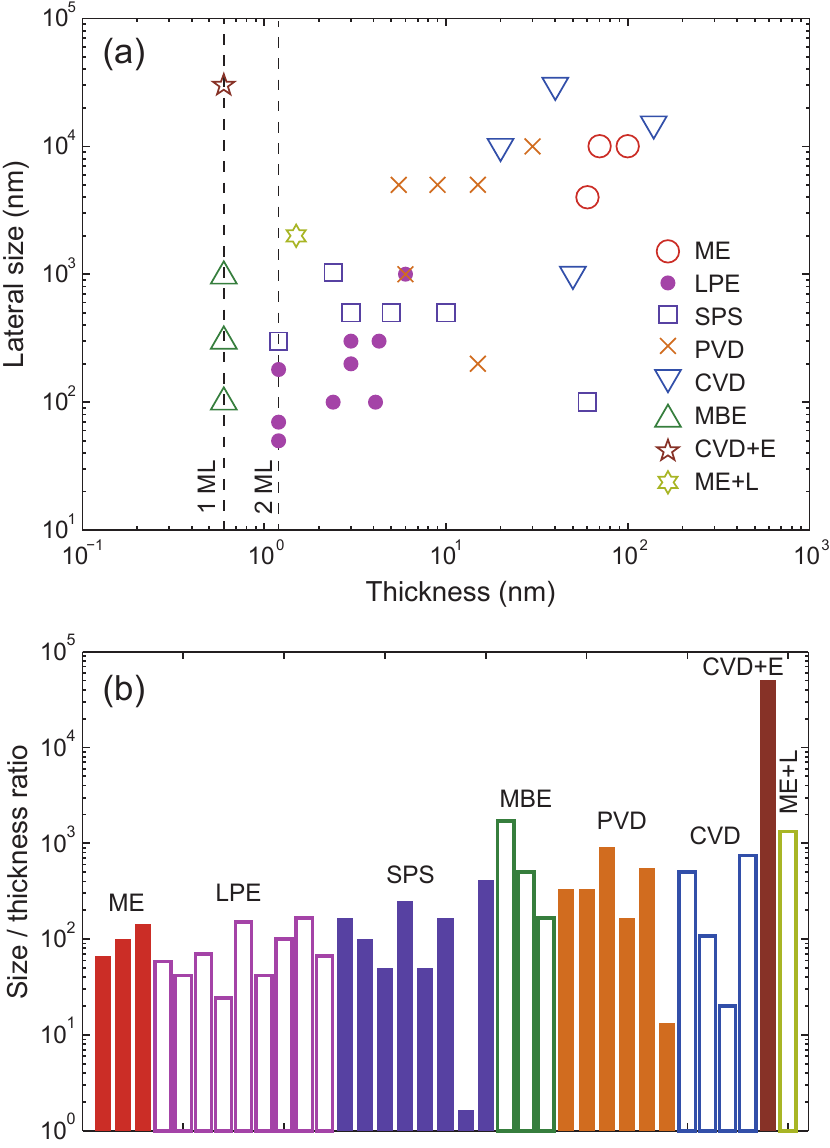}
\caption{(a) Plot of lateral size against the lowest thickness of MX flakes fabricated using various methods, from the references given in Table~\ref{results_summary}. Abbreviations, SPS: solution phase synthesis; CVD+E: CVD + nitrogen etching; ME+L: mechanical exfoliation + laser etching. (b) Comparison of the ratio between the lateral size and thickness of the MX flakes in (a).}\label{stat}
\end{figure}

{In Fig.~\ref{stat}, we summarize the thickness and lateral size of MX flakes created by different methods listed in Table~\ref{results_summary}.} In general, methods that can generate thin and large flakes are favored, which corresponds to the upper-left corner of Fig.~\ref{stat}(a). However, it is clearly shown that, except from flakes prepared by etching methods, the dimensions of all the as-prepared flakes, no matter whether exfoliated or synthesized, follow a simple rule: the thickness and lateral size are positively correlated. On the one hand, mechanical exfoliation, PVD and CVD methods can produce flakes as large as tens of $\mu$m in extent, but it is difficult to reduce the thickness below 5 nm. On the other hand, MBE, liquid phase exfoliation and solution phase synthesis methods can create atomically-thin, or even monolayer thick flakes, but the average lateral sizes of all the flakes thinner than 4 MLs are smaller than 1 $\mu$m. Again, the relatively large inter-layer coupling energy in MX materials is probably the reason for this dependence.

Based on current studies, there are two possible routes to solve this dilemma. First, since this limit originates from the intrinsic properties of MX, one can introduce extrinsic factors to enhance the anisotropy of MX flakes. Proper choice of substrates and synthesis recipes could make the tendency for lateral growth stronger. For example, monolayer nanosheets of SnTe, over 1 $\mu$m in extent, have been found using MBE growth. \cite{Chang16_Science_SnTe} Second, chemical and physical etching methods can overcome the intrinsic limit due to the material's anisotropy, and thus are promising for creating large monolayer MX flakes that are suitable for device fabrication. Especially, the self-limiting etching mechanism, which has been observed both in nitrogen and laser etching, guarantees a uniform final thickness. Such a self-limiting mechanism is likely to be a result of the strong interaction at the interface between the substrate and the MX flakes, which, in turn, leads to new questions: how the substrate influences the atomic and electronic structures of MX flakes, and whether the monolayer MX flakes strongly bond to the substrate while still {maintaining} their intrinsic properties, are important topics that should be explored in future studies.

These points above are better supported by exploring the {anisotropy} ratio, which is the ratio between the lateral size and the thickness of the MX flakes created by different methods, in Fig.~\ref{stat}(b). For the methods that do not introduce any extra anisotropy, such as exfoliation and solution phase synthesis, the {anisotropy} ratios are generally around $10^2$. For the methods that require substrates, which introduces extra anisotropy to the synthesis process, such as MBE, PVD and CVD, the {anisotropy} ratio can be close to $10^3$. With etching methods, the anisotropic ratio can be further pushed above $10^4$, where the lateral size is only limited by that of the pristine thicker flakes.

\section{Conclusions}\label{sec5}

In this Perspective, we have reviewed experimental approaches that have successfully created monolayer MX, as well as those that could be used to generate monolayer MX in the near future. Currently, clear reports of the creation of MX monolayers include MBE grown SnTe, PbTe and SnSe (among which the growth of monolayer SnSe adopts a two-step recipe), and SnSe flakes that are synthesized by CVD and then etched in a nitrogen atmosphere. Laser etching of mechanically exfoliated GeSe nanoflakes can also achieve a stable thickness of 1.5 nm. {Despite them being claimed as monolayers}, the actual thickness of the laser thinned samples needs to be cross-checked by further experiments. Both the chemical nitrogen etching and physical laser etching processes exhibit self-limiting mechanisms, which allow large windows for the etching parameters. Furthermore, 2-ML thick MX flakes can be created through liquid phase exfoliation and solution phase synthesis. We have quantitatively compared the dimensions of the MX flakes fabricated in various approaches, and suggest that the introduction of extra anisotropy during synthesis or by using post-etching techniques are the keys to the creation of {large-size} monolayer MX flakes.

{Note added: We noticed that a study of the PVD growth of monolayer SnS and the demonstration of its in-plane ferroelectricity was published during peer review. \cite{Higashitarumizu20_NC_SnS}}

\begin{acknowledgments}
K.C. was supported by the startup funding from Beijing Academy of Quantum Information Sciences. K.C. and S.S.P.P. were funded by Deutsche Forschungsgemeinschaft (DFG, German Research Foundation)--Project No. PA 1812/2-1. The original data that have not been published elsewhere are available from the corresponding author upon reasonable request.
\end{acknowledgments}


%

\end{document}